\documentclass[prd, preprint, longbibliography, 12pt]{revtex4-1}

\usepackage{amsmath}
\usepackage{amssymb}
\usepackage{setspace}
\usepackage{graphicx}
\usepackage{natbib}
\usepackage{float}
\usepackage{tensor}
\usepackage[utf8]{inputenc}
\usepackage{amsfonts}
\usepackage{braket}
\usepackage{esint}
\usepackage{breqn}
\usepackage{IEEEtrantools}

\begin{document}
 
 %

\begin{center}
{ \large \bf Spontaneous Quantum Gravity
 }


\vskip 0.2 in

{\large{\bf Tejinder P.  Singh }}

\medskip

{\it Tata Institute of Fundamental Research,}
{\it Homi Bhabha Road, Mumbai 400005, India}\\
\bigskip
 {\tt tpsingh@tifr.res.in}

\end{center}

\centerline{\bf ABSTRACT}
\noindent  This article gives  an elementary account of the recently proposed theory of spontaneous quantum gravity. It is argued that a viable quantum theory of gravity should be falsifiable, and hence it should dynamically explain the observed absence of quantum superpositions of space-time geometries in its classical limit.
\bigskip

\bigskip

\bigskip

\section{Schrodinger's cat in the 21st century}
If a physical theory agrees with some of the data, but not with all the data, it should be replaced by a falsifiable new theory which agrees with all of the data. The new theory should reduce to the old theory in those domains where the old theory agrees with the data.

Such a strategy worked successfully in the transition from Newtonian mechanics to special relativity, and it worked successfully in the transition from Newtonian gravitation to Einstein's general relativity. We can say that special relativity is a cover for Newton's mechanics, and general relativity is a cover for Newtonian gravitation. But the transition from Newtonian mechanics to quantum mechanics is a different story altogether!! Quantum mechanics is not a cover for classical mechanics.

Quantum mechanics was invented to explain data such as the black-body radiation spectrum, atomic spectra, and the photoelectric effect, which Newton's mechanics fails to explain. And quantum theory explains these, and much much more, beautifully. But quantum theory fails to explain the data that Newton's mechanics explains! So we need a new theory which will agree with both quantum mechanics, as well as with classical mechanics.

At the heart of the disagreement between classical and quantum mechanics is the very elegant quantum linear superposition principle, which says that if a quantum system can be in State $A$ and if it can be in State $B$, then it can also be in the superposed state $A+B$. In particular, the superposition principle is known to hold for positions of a particle. If an electron can be here, and if the same electron can be there, it can also simultaneously be here as well as there. This is how we understand the appearance of interference fringes on the screen in a double-slit experiment with electrons.

The quantum superposition principle has been experimentally verified to hold for photons, neutrons, atoms, small molecules, and is known to hold for objects as heavy as 25,000 a.m.u. That is, an object made of  25,000 nucleons. Experimentalists would love to test the principle for even heavier objects, but it is technologically extremely challenging. They are at it.

But the superposition principle fails for large objects that we see in our day to day life, obviously. We never see a chair to be here and there at the same time. Nor do we see a planet to be in the north and in the south at the same time. Yet the motions of chairs and planets is successfully described by Newton's mechanics. This is what we mean when we say that quantum mechanics fails for large objects and disagrees with classical mechanics. In fact, Schrodinger's equation predicts that a chair can be here and there simultaneously. The mass of the object described by the Schrodinger equation is completely arbitrary. This mass can be as large as we please. Hence Schrodinger's equation should hold for a chair, and superposition should have been observed, but it is not observed. Another way to appreciate the problem is that the chair is made of elementary particles which themselves obey the superposition principle. Why is it that when we put many such particles together, superposition breaks down?

What is the way out of this contradiction between quantum mechanics and Newtonian mechanics? We need a cover for Newton's mechanics, which will agree with quantum mechanics for small objects. Such a new theory was proposed by physicists Ghirardi, Rimini, Weber and Pearle (GRWP)  in the 1980s. Their idea is beautiful and simple. They said, look - quantum mechanics says that a superposition, once created, lasts forever. This is because the Schrodinger equation is linear and deterministic. On the other hand, Newton's mechanics does not allow for position superpositions at all. GRWP said, let us make a very very small modification to quantum theory. Let us propose that superposition of two position states of a particle, say a proton, does not last forever, but lasts for an extremely long time $T$. That is, the mean superposition life-time of two position states of a proton is not infinite, but a large number $T$. For definiteness, they proposed $T$ to have the value $10^{17}$ seconds, (which also happens to be the age of the universe) for a nucleon. After a time $T$, the superposition is assumed to spontaneously collapse to one or the other states which were superposed (here or there). This is the GRWP theory of spontaneous collapse. Remember, $T$ is the mean lifetime, and collapse is a random (Poisson) process in time. There is always a tiny probability for spontaneous collapse to take place in a time much smaller than $T$.

This small modification to quantum theory suffices to provide us with a new theory which reduces to Newton's mechanics for large objects. Again, this works in a very elegant manner. Consider, for a start, a deutron - the nucleus of the deutirium atom, which is a bound state [also an entangled state] of a neutron and a proton. Now, for a deutron to undergo a spontaneous collapse, starting from a superposed state, it is enough for either the proton to undergo a spontaneous collapse, or for the neutron to undergo spontaneous collapse. One particle will take the other with it, because they are bound (entangled). You can then reason that  the superposition lifetime for the deutron is halved, it is $T/2$, because there are two independent ways in which the collapse can happen.

There, now you have it. A large object such as a chair is made is of an enormous number of nucleons and electrons. If there are $N$ particles in the chair, the chair can be in a superposed state (here and there) only for a time $T/N$. [Because any one particle collapsing will take the whole chair with it].  But $N$ for a chair is huge, say $10{23}$.. Since $T$ is assumed to be $10^{17}$ seconds, $T/N$ is a mere millionth of a second. The superposed state for a chair lasts for such a short time, that we do not even notice it. Schrodinger's cat is dead as well as alive for a millionth of a second; after that it is dead, or alive. That is why large objects appear to obey Newton's mechanics.

In this way, the theory of spontaneous collapse is the cover for Newton's mechanics. The cover theory reduces to quantum mechanics for small objects. This is because for small objects, the superposition life-time is so enormous as to be practically infinite, as demanded by quantum mechanics.

The GRWP theory would occupy the same place of pride as special relativity and general relativity, if it were to be confirmed by experiment. Experimentalists are working hard to test it. The current experimental bound on $T$ is that $T > 10^8$ seconds. Recall that GRWP say that $T=10^{17}$ seconds. and quantum mechanics says that $T$ is infinite. Still nine more orders of magnitude to go before GRWP is ruled out. Note that a confirmed detection of spontaneous collapse below GRWP value will also prove the theory of spontaneous collapse. The theory will be ruled out if experiments will push the bound on $T$ beyond the GRWP value. If $T$ is higher than the GRWP value, then for large objects $T/N$ will approach time scales larger than a millionth of a second,  which means we would see a chair here and there at the same time. Thus, values of $T$ larger than the GRWP value do not provide a cover theory for Newton's mechanics \cite{RMP:2012}.

\section{From quantum foundations to quantum gravity}

Quantum theory was invented to explain experimental data which could not be explained by Newton’s mechanics. There is no such clear-cut compelling observational evidence to suggest that gravity must be quantised. It could be said that the classical general theory of relativity agrees with every experiment / observation carried out till date. It may or may not turn out to be the case that understanding dark energy, cosmological constant, dark matter, require us to unify quantum and gravity. It may or may not turn out to be the case that the gravitational singularities that arise in general relativity require us to quantise the theory.

However, there are reasons within quantum (field) theory which compel us to consider a non-classical description of space-time and of space-time geometry. Quantum theory needs a time parameter, so as to describe the evolution of quantum systems. This time parameter is a part of a classical space-time, whose geometry is determined by classical bodies according to the laws of general relativity. But classical bodies are a limiting case of quantum systems. It should be possible to describe the dynamics of a quantum system without having any dependence [direct or indirect] on classical bodies. And yet, in the absence of classical bodies [i.e. if all matter were quantum], one cannot have a classical space-time geometry, nor a classical space-time manifold. This is a consequence of the so-called Einstein hole argument, which you can learn more about, from say this video: The problem of time in quantum theory [https://www.youtube.com/watch?v=fGdOTokept8].

Thus, we must have a formulation of quantum theory which does not depend on classical space-time. This will be our sought for quantum theory of gravity. We do not quantise gravity or space-time. Rather, we remove space-time from quantum (field) theory.

Can this goal be achieved by applying the rules of quantum theory to a classical theory of gravity? The answer is no. Firstly, the quantum rules are written down assuming classical time to exist. How then can we apply these rules to quantise the very time parameter whose classical existence was in the first place assumed, for writing these rules? There is no guarantee that this [admittedly illogical] step will lead us to the correct theory.

But secondly, there is an even more serious reason for the answer to be no. A classical theory of gravity does not permit superposition of space-time geometries: such superpositions are never observed, just as a chair is never observed in more than one place at the same time. On the other hand, a quantum gravity theory resulting from quantising classical gravity will naturally admit superpositions of geometries. And the theory will predict a superposition of geometries even when the bodies producing these geometries become large and classical. Same way as quantum theory predicts that a chair can be here and there at the same time. This is the Schrodinger cat paradox in the context of spacetime geometries. In the language of the previous section, such a quantum gravity theory is not the cover of classical general relativity.

To recover classical general relativity from quantum gravity, the sought for quantum gravity theory must admit a spontaneous collapse of superposed geometries, precisely in the spirit of the GRWP theory discussed in the previous post. Let us name such quantum gravity, which admits spontaneous collapse of geometry, as spontaneous quantum gravity. From here it is easy to reason that the absence of macroscopic position superpositions in the classical world  is a consequence of spontaneous quantum gravity. Classical space emerges from quantum gravity, and moreover for classical space to exist, macroscopic bodies must be classical (not quantum). Thus the GRWP theory is a consequence of quantum gravity. This is readily seen in another way. Imagine a situation in which no quantum object has yet undergone spontaneous collapse: then there is only quantum matter and quantum space-time - the domain of quantum gravity. It follows that GRWP must arise from quantum gravity. Hence the name spontaneous quantum gravity.

Spontaneous quantum gravity [SQG] is the cover for classical general relativity, same way as GRWP is cover for Newton’s mechanics. SQG is falsifiable, because it predicts spontaneous collapse, and the latter is falsifiable. Recall that in GRWP, spontaneous collapse is proposed in an ad hoc manner. But in SQG, spontaneous collapse is not ad hoc. It is a consequence of the structure and dynamics of the theory.

One could well ask, quantum (field) theories of other interactions, such as QED, are not covers of their classical counterparts, such as Maxwell’s electrodynamics. Yet, why is QED such a successful theory, even though it does not explain the absence of superpositions in the classical electro-magnetic world? The answer is that QED is not a quantum theory of spacetime. It is the quantum theory of a field which lives on spacetime, and of the electric charges which produce these fields. Quantum gravity has to explain how classical space emerges, and since classical space is tied to absence of position superpositions in macroscopic bodies, quantum gravity has to explain why macroscopic bodies are classical. Once the position of macroscopic bodies is localised, their mass is localised, and their electric charge is localised too, and hence the associated electromagnetic fields are classical. Electromagnetic fields live on classical space-time, and require space-time to pre-exist. Space-time does not live on a classical electromagnetic field! Hence the buck stops with gravity.

We saw in the previous section that GRWP theory is the cover for Newton’s mechanics, and for small systems GRWP reduces to quantum theory, because the rate of spontaneous localisation is negligible for small systems. In the present section we see that spontaneous quantum gravity is the cover for classical general relativity, and for small objects it reduces to …? Reduces to what? We expect it to reduce to quantum gravity, because now the rate of spontaneous collapse of geometries is negligible, where by quantum gravity we mean quantisation of classical general relativity. [Incidentally, when we talk of rate of collapse of superposed space-time geometries, how is rate defined? What is this time parameter which keeps the rate? We will take up this deep question subsequently]. Thus, in all likelihood, we expect that the limit of SQG for small objects is related to loop quantum gravity. So we can say:

\centerline{GRWP theory = Quantum theory + Spontaneous collapse}

\centerline{SQG = Quantum gravity + Spontaneous collapse}

The GRWP theory already exists and is well defined and is being tested in the laboratory. How do we mathematically formulate spontaneous quantum gravity? We will take this up in a section below.

\section{The quantum measurement problem, and its solution via spontaneous collapse theory}

A non-relativistic quantum system evolves according to the Schrodinger equation, which is linear and deterministic.  Given the initial state, the final state can be precisely determined, by solving the Schrodinger equation. And if the initial state is a superposition of two eigenstates of some observable, it will evolve into a final state which is also a superposition of those two eigenstates. Moreover, if the equation is being considered for an object of some mass, the value of the mass can be arbitrarily large, according to the Schrodinger equation.

Keeping all this in mind, let us consider what happens when a quantum system, being described by the Schrodinger equation, meets a measuring apparatus. Say the apparatus measures spin of an incoming quantum particle, which is in a linear superposition of two states, say (spin up), and (spin down), having a complex amplitude $A$ to be in up state, and amplitude $B$ to be in down state. Suppose the spin value is deduced from the position of a pointer, and the pointer points up if the state is spin-up, and the pointer points down if the state is spin-down. According to the Schrodinger equation, if the particle is in a superposition, the following state should be observed after the measurement has been done:

\centerline{A (spin up) (pointer up ) + B (spin down) (pointer down)}
\noindent [Such a state is called an entangled state. The quantum particle and the apparatus have become `entangled’ after interacting]. The superposed state  of the particle should force the pointer also into a superposed state. However, what is actually seen after the measurement is something completely different, and extremely surprising.

After the measurement the quantum particle is found either in up state with pointer pointing up, or in down state with pointer pointing down. Which of the two? Its random. The outcome is random, and cannot be predicted beforehand. If one does the same experiment many many times, sometimes the outcome is up, sometimes it is down. But the fraction of times the outcome is up, is experimentally found to be given by $|A|^2$, the square modulus of $A$. The fraction of times the outcome is down, is given by $|B|^2$. This is the so-called Born probability rule - an empirical rule always found to hold in quantum measurements, namely that the probability of an outcome is given by the square modulus of the corresponding amplitude.

As you can see, what actually happens during a measurement completely disagrees with the Schrodinger equation. Quantum mechanics fails during the measurement process. It fails on the following counts: (i) Superposition is lost, whereas it should not have been lost. Why did `up + down’ go to either up or down, even though the equation is linear? (ii) The Schrodinger equation is deterministic. Why then are the outcomes random and unpredictable? (iii) Since the equation is deterministic, it has nothing to do with probabilities! Where have the probabilities arisen from?  Why does the Born rule hold - it is an experimentally observed rule, which obviously cannot be derived from the Schrodinger equation. And why do the probabilities mysteriously depend on the amplitudes $A$ and $B$, when probability has nothing to do with Schrodinger equation, whereas $A$ and $B$ are properties of the state that evolves according to the Schrodinger equation? This set of disagreements between theory and experiment is commonly referred to as the quantum measurement problem.

As we said at the beginning of the first section, if a theory agrees with some data, but not with all data, it should be replaced by a new theory which agrees with all of the data. The Schrodinger equation correctly describes the experimentally observed motion of the particle before it meets the measuring apparatus, but fails to describe what happened during the measurement process. Hence, it should be replaced by a new equation which agrees with the Schrodinger equation before the measurement (i.e. when only the quantum particle is being considered). But the new equation should disagree with the Schrodinger equation during the measurement process, in such a way that the new equation resolves the three counts on which the Schrodinger equation fails, and explains what actually happens during a measurement.

The GRWP theory of spontaneous collapse provides precisely the correct new equation for this purpose. As we know, for the quantum particle, GRWP agrees very well with quantum mechanics and with the  Schrodinger equation, because the superposition life-time (being of the order $T$) is so large as to be practically infinite. However, now let us apply GRWP to the measurement process, and in particular to the entangled state written above. This is a superposition of two position states of the pointer (up and down). But the pointer is made of enormously many particles, and we know according to GRWP that the superposed state for such a large object is extremely short-lived, lasting only for a millionth of a second or so. After this much time the pointer spontaneously and randomly collapses to the up state or to the down state, and takes the spinning quantum particle with it. So we infer that the spin state of the particle has randomly changed to spin-up or spin-down. This is how GRWP theory solves the quantum measurement problem.

But what about the Born probability rule? Does the GRWP theory prove this modulus-square rule? Alas, it does not. It takes the rule as an assumption, a given property of spontaneous collapse. In subsequent sections, we will see how the Born rule arises as a consequence of spontaneous quantum gravity. Note that from the point of view of the GRWP theory, there is nothing special about a quantum measurement. It is just a particular case of macroscopic position superposition, which according to GRWP is short-lived. The measurement problem is the same as the problem of macroscopic superpositions not being observed in nature, so the proposed solution to the two problems is also the same, i.e. the falsifiable GRWP theory.

\section{Limitations of the spontaneous collapse theory}

The GRWP theory of spontaneous localisation is a falsifiable phenomenological theory. It is designed to provide a dynamical solution to the quantum measurement problem, and to provide a cover for Newtonian mechanics that agrees with quantum mechanics for microscopic systems. The theory is precise enough for experimentalists to be able to test it, and confirm it or rule it out. That is why experimentalists are testing it. For knowing about some of the latest experimental developments on this front, the reader can visit tequantum.eu The most direct way to test GRWP is to verify if the principle of linear superposition holds for large objects. As we saw before, quantum mechanics predicts that a superposition of two position states of an object lasts forever. On the other hand GRWP predict that the superposition lasts for a time $T/N$, with $N$ being the number of particles in the superposed object. So experimentalists prepare a superposed state, say by using a diffraction grating, and watch if it decays during the time of observation. If it does not, quantum mechanics wins, and one puts a lower bound on $T$. These are the so-called interferometric tests of (spontaneous) collapse models.

However, in recent years, the so-called non-interferometric tests of collapse models have moved centre-stage. Every time an object in a superposed state undergoes spontaneous collapse to some random location in space, its wave function expands again, and then again it collapses, with the mean life-time between collapses being $T/N$. These repeated random collapses amount to a random walk, with which is associated a tiny amount of kinetic energy. Spontaneous collapses cause the quantum object to gain a very tiny amount of energy. After cooling the object to extremely low temperature - few milli-Kelvins - and low pressure, one can attempt to look for this random walk, which obviously is in violation of quantum mechanics. Such experiments are currently in an exciting stage, and we might hear of some exciting results in the next few years.

The GRWP theory mathematically amounts to a modification of the Schrodinger equation. One adds a non-Hermitian (random) term to the Hamiltonian of the quantum system. Random because we want the resulting spontaneous collapse to result randomly. Non-Hermitean because we want one of the superposed states to grow exponentially, and the other one to decay exponentially (i.e. destroy superposition). Now, adding such a term implies that evolution no longer preserves norm of the quantum state. However, if the Born probability rule has to be obtained, norm must be preserved. Thus, a new quantum state is defined, by scaling with the norm of the old state. The new state now obeys a non-linear and non-Hermitian  stochastic differential equation, which describes spontaneous collapse theory. The equation has the standard linear and Hermitean part which describes Schrodinger evolution, and in addition it has a non-linear, non-Hermitian part which describes non-unitary evolution, which causes spontaneous localisation and breaks position superposition. It is this equation which the experimentalists are testing. For microscopic objects, the predictions of this equation are extremely close to that of the Schrodinger equation (the non-linear part is negligible), but for macroscopic objects the deviations from quantum mechanics become significant. Here, the predictions of the new equation differ from those of quantum mechanics, and are falsifiable.

A theorist can raise a whole lot of questions and criticisms against the theory of spontaneous collapse, and these need to be addressed and resolved, so that the theory becomes more credible. As it stands, the theory is ad hoc in various ways. What is the origin of the random noise which has been added to the Schrodinger equation? What is the spectrum of this noise? Why should the collapse rate parameter $T$ have this particular value of $10^{17}$ sec, and no other value? What causes spontaneous collapse in the first place?  Why should the norm of the state vector be preserved, in spite of the evolution being non-unitary?

Perhaps the most serious criticism against collapse models is that they are non-relativistic. And attempts to  make a  relativistic Lorentz-invariant theory of spontaneous collapse have not been successful. Now, our most successful physical theories are relativistic quantum field theories, which describe the standard model of particle physics, and show excellent agreement with experiments. The Schrodinger equation is readily shown to be the non-relativistic approximation to the Dirac equation, which is relativistic. How then are we adapt the stochastic corrections provided by GRWP to the context of a quantum field theory?

In the opinion of this author, there is a convincing reason why one cannot have a relativistic theory of collapse, without making additional conceptual changes. Recall that spontaneous collapse takes place in position space: the position operator of a particle jumps to a specific eigenvalue, causing spontaneous localisation. Now, in special relativity, we expect position and time to be treated in a symmetric fashion. Hence, in order to make a relativistic theory of spontaneous collapse, we must allow also for spontaneous localisation in time! For that to happen, time will have to be treated as an operator, just like position is an operator in quantum mechanics. In that case, time loses its role as a parameter for defining evolution, and we are then compelled to introduce into relativistic quantum mechanics a new absolute and universal time parameter,  which can be used to define evolution. To summarise, in order to have a relativistic theory of spontaneous localisation, space-time coordinates must be turned into space-time operators in quantum theory, which can undergo spontaneous collapse, and time evolution has to be described by a new absolute time parameter.

There is a lot that has been said and encoded in the previous paragraph, so we now  dwell carefully on the various issues that arise. Firstly, why is it that relativistic spontaneous collapse forces us to treat ordinary time as an operator, whereas no such compulsion arises in standard relativistic quantum field theory? The answer is subtle. So long as spontaneous collapse in position can be ignored [as of course is the case for QFT], spontaneous collapse in time can be ignored as well, and we have our Lorentz invariant quantum field theory. In non-relativistic quantum mechanics, switching on collapse in position space does not compel us to switch on collapse in time space. Because the theory is Galilean invariant; it is not Lorentz invariant, and time is absolute. On the other hand, in the relativistic case, Lorentz invariance compels us to introduce spontaneous collapse in time, soon as we introduce spontaneous collapse in position. In turn, that forces us to introduce an absolute universal time parameter.

It should be mentioned though, that such a (covariant) formulation of relativistic quantum field theory, which treats position as well as time as operators, does exist. It is known as the Horwitz-Stueckelberg theory and the reader can read more about in Lawrence Horwitz’s book `Relativistic Quantum Mechanics’ [Springer, 2015]. The book also discusses the phenomenon of `quantum interference of time’ which will inevitably arise once time has been made an operator. It means that a quantum particle can be at more than one time, at a given universal time. Sone researchers claim that experimental evidence for quantum interference of time already exists. In any case, it is of great importance to perform experiments to look for quantum interference of time. Spatial quantum interference is comparatively much easier to detect, but as and when quantum interference of time is detected, QFT and relativistic quantum mechanics will have to be written in the language of the Horwitz-Stueckelberg theory.

Spontaneous collapse in time may appear to be a bizarre phenomenon, but we have been led to it in a logical inescapable manner. In order to have a cover theory of Newton mechanics which agrees with quantum mechanics for microscopic systems, we are compelled to introduce spontaneous collapse in position. In order to make this collapse theory relativistic, we are compelled to introduce spontaneous collapse in time. Experimentalists ought to look for collapse in time, just as they are testing the GRWP theory.

We can also ask: just as a chair is never found in two places at the same time, why is the chair never found in two times at the same place?! This maybe attributed to rapid spontaneous collapse in time, caused by the chair being made of many many particles.  Spontaneous collapse in space as well as time together define classical events. We expect a quantum particle such as an electron to be at more than one time at the same place (as already hinted at by the path integral formulation of relativistic quantum mechanics) in a very real and physical sense. A quantum particle senses the past as well as the future `simultaneously’. What implications does this have for our understanding of physical reality?

Lastly we mention that the universal absolute time parameter which relativistic collapse theories compel us to introduce, turns out to be rooted in non-commutative geometry, and  in the theory of spontaneous quantum gravity, which we will take up in subsequent sections. But we can already see that the need to introduce space-time coordinate operators already takes us towards quantum gravity, and away from classical space-time geometries. And later we will see how and why spontaneous collapse is an inevitable consequence of spontaneous quantum gravity. The ad hoc nature of the GRWP theory is removed then, because it emerges from an underlying physical theory.

Because spontaneous collapse is essential for localisation of macroscopic objects and resolution of the quantum measurement problem, and because localisation of macroscopic objects is essential for the existence of space-time [Einstein hole argument] and because space-time emerges from quantum gravity, we conclude that the solution of the quantum measurement problem comes from a quantum theory of gravity. Thus one cannot construct a quantum theory of gravity by quantising classical gravity, because doing so does not give a quantum theory of gravity which will dynamically explain absence of superposition of classical space-time geometries. In subsequent sections, we will clearly see how and why Planck length appears in the stochastic part of the non-linear Schrodinger equation which explains spontaneous collapse.

\section{The origin of spontaneous localisation: trace Dynamics (I)}

We have seen earlier that spontaneous localisation is a falsifiable but ad hoc modification of the Schrodinger equation, for explaining the absence of macroscopic position superpositions. One would like to derive this collapse theory from an underlying dynamics, based on a symmetry principle. One such theory, known as Trace Dynamics (TD) has been developed by Stephen Adler (IAS, Princeton) and collaborators. The best source to read about this theory is Adler’s book `Quantum theory as an emergent phenomenon’ [Cambridge University Press, 2004].

Put in brief, Trace Dynamics is the dynamics of matrix models which obey a global unitary invariance. Quantum (field) theory is emergent as the statistical mechanics of these matrix models. TD is assumed to operate at the Planck scale, although space-time is assumed to be Minkowski space-time. The matrix models describe the dynamics of fermionic matter, as well as of gauge fields, although no specific form of the Lagrangian of the theory is prescribed. Gravity is not included.

What is the motivation behind trace dynamics? One would not like to arrive at quantum theory by quantising a classical theory. This is considered unsatisfactory because the classical theory is only a limiting case of quantum theory - one should not have to know the limit of a theory to construct the theory; it should be the other way around. [We do not construct special / general relativity by `relativising’ Newtonian mechanics / gravitation. The relativity theories are built from their own symmetry principles, and yield the Newtonian theories in the limit]. Trace dynamics is a first principles theory, from which quantum, and classical mechanics, are emergent as approximations.

In TD, particles and fields are not described by real numbers, but by matrices (equivalently operators). Consider for instance the Newtonian Lagrangian dynamics of a collection of particles with configuration variables $q_i$. From these variables and their time derivatives we can make the Lagrangian, and from there obtain the equations of motion. To construct trace dynamics, assume instead that each of the $q_i$ is a matrix/operator. We will construct its corresponding velocity by taking the time derivative of the matrix, which is equivalent to taking time derivative of each matrix element. Given this set of configuration matrices and their velocities, the Lagrangian of TD is constructed by making a polynomial from matrix products, and then taking the matrix trace of this polynomial. Thus the Lagrangian is a scalar (a real number), rightfully called the trace Lagrangian. The action in TD is the time integral of this trace Lagrangian. Lagrange equations of motion are obtained by extremising the action by varying it with respect to the operator configuration variables. This requires the introduction of a `trace derivative’ - which is a natural method for differentiating  trace of a polynomial w.r.t. an operator. The resulting Lagrange equations are operator equations. Also, TD is constructed to be a Lorentz invariant theory. Importantly, the configuration variables and their conjugate momenta, all non-commute with each other. Unlike in quantum theory, the commutators in TD are arbitrary and time-dependent, in general.

What, one might ask, is the point of these matrices/operators? What is the physical interpretation of their matrix elements? TD is a pre-quantum theory, more general than quantum theory, operating at the Planck scale; a theory from which quantum mechanics will be derived as an emergent approximation, at energy scales below the Planck scale. The eigenvalues and eigenstates of these matrices represent possible values that these degrees of freedom can take, and superposition  is possible too; yet the rules of evolution are not those of quantum theory.

Trace dynamics can be extended to fields too, by dividing three-space into cells and assigning one q-operator for every cell, which then represents the field value in that cell. Alternatively, one can take the continuum limit of the $N$-particle trace dynamics.

The elements of the matrices in TD are made from complex numbers and complex Grassmann numbers. A complex Grassmann number is made from two real Grassmann numbers. A real Grassmann number is made from products of Grassmann elements. Grassmann elements anti-commute with each other, unlike ordinary numbers which commute with each other. Thus the square of a Grassmann number is zero. A product of an even number of Grassmann elements commutes with every Grassmann element, and together with unity these form an even-grade Grassmann algebra, which is used to represent bosonic fields. A product of an odd number of Grassmann elements anti -commutes with other odd number products of Grassmann elements: together these form the odd-grade sector of the Grassmann algebra. These are used to represent fermionic fields. A general matrix made from complex Grassmann numbers can be written as a sum of two matrices, one made from even-grade elements (and called bosonic) and one made from odd-grade (and called fermionic).

In matrix dynamics the trace Hamiltonian of the system is conserved, as can be expected. In addition though, trace dynamics possesses a fascinating conserved charge, which results from the global unitary invariance of the trace Lagrangian and trace Hamiltonian. This charge is known as the Adler-Millard charge, it has no analog in ordinary classical dynamics, and is responsible for the emergence of quantum theory from trace dynamics. It is given by the sum (over bosonic degrees of freedom) of the commutators $[q,p]$, minus the sum (over fermionic degrees of freedom) of the anti-commutators $\{q,p\}$. Note that this charge has the dimensions of action. If the Hamiltonian  is self-adjoint, the AM charge is anti-self-adjoint. If the Hamiltonian were to also have an anti-self-adjoiint piece, the AM charge picks up a self-adjoint component. One can also define a generalised Poisson bracket in trace dynamics, and express the dynamics as Hamilton’s equations of motion, and also in terms of Poisson brackets. In general, evolution in trace dynamics is not unitary, and differs from the evolution given by the Heisenberg equations of motion in quantum theory.

Trace dynamics in many ways resembles matrix models that have been studied in quantum field theory. However, in matrix models, canonical quantum commutation relations are imposed on the matrix elements, thus leading to the standard quantisation. Unlike matrix models though, one does not quantise a trace dynamics. Instead one asks, what does the coarse grained dynamics look like, if trace dynamics is averaged over time-intervals much longer than Planck time, and one is interested in the emergent dynamics at energy scales much smaller than Planck scale? It turns out that this emergent dynamics is quantum dynamics. This is established by applying the conventional techniques of statistical mechanics to trace dynamics, a topic which we will take up in the next section.

\section{The origin of spontaneous localisation: trace Dynamics (II)}
As we saw earlier, trace dynamics is a matrix dynamics with a global unitary invariance, which operates at the Planck scale. Now we ask this question: suppose we do not wish to examine the dynamics on time scales of the order of Planck time or even smaller times, and we coarse-grain the trace dynamics over time steps much larger than Planck times, what is this time-averaged dynamics? It is shown, using the techniques of statistical mechanics, that this averaged dynamics is quantum (field) theory. No specific fine-tuning is done in the underlying trace dynamics so as to ensure that the emergent dynamics is quantum dynamics. Of central importance is the fact that as a  consequence of the global unitary invariance of the trace Lagrangian, trace dynamics possesses a conserved charge, the Adler-Millard charge. This charge has dimensions of action, and its equipartition as a result of doing the statistical mechanics is what is responsible for the emergence of quantum theory. It is plausible to assume that since the emergent theory results from averaging over time intervals much larger than Planck time, the energy scale associated with the emergent quantum system is much smaller than Planck energy. And this is of course true for the systems we currently study in laboratories while applying quantum field theory to the standard model of particle physics. It is already implicit in this analysis above that the dynamical laws on the Planck scale are not those of quantum field theory, but those of trace dynamics. This proposal will become much more plausible when we incorporate gravity into trace dynamics, in the theory of Spontaneous Quantum Gravity.

The principles of statistical mechanics are generally employed to derive the properties and thermodynamic laws for macroscopic systems, starting from the laws of atomic theory. Central to these principles is the fact that a macroscopic system is made of an enormously many constituent particles, whose individual motions we know how to describe, but we are not interested in. The macro-state is then determined by maximising the entropy made from all the microstates consistent with the values of physical attributes describing the macro-state (say constant temperature, or constant energy).

In the present instance of trace dynamics, these principles of statistical mechanics are being put to a different use, not for finding what happens to trace dynamics when we are considering a macroscopic system. Instead, we want to know what is the average dynamics of a system of one or more matrices obeying trace dynamics, when the dynamics is coarse-grained over time scales larger than Planck time scale. To find this average dynamics, we consider an ensemble of a very large number of matrices, each of which obeys trace dynamics at the Planck scale. Each one of them follows a different trajectory in the phase space, but we can assume [the ergodic hypothesis] that the ensemble average of the dynamics at any one time represents the long time average (i.e. the averaged dynamics of a matrix when coarse-grained over intervals larger than Planck time). We want to know the equilibrium ensemble, subject to the constancy of the trace Hamiltonian and the Adler-Millard charge.

One starts by defining a volume measure in the phase space made from the matrix elements - if there are $N$ matrices in the trace dynamics, there is one canonical pair in the phase space for each of the elements of every one of the $N$ matrices. A Liouville theorem is proved, namely that trace dynamics evolution preserves a volume measure in phase space. Next, the equilibrium density  distribution function in phase space is defined, which as usual gives the probability of finding the system in a given infinitesimal phase space volume. It is also shown that the ensemble average of the AM charge is a constant times a unit matrix, and since this constant has dimensions of action, it will eventually be identified with Planck’s constant subsequent to the emergence of the quantum dynamics.

The Boltzmann entropy is defined from the density distribution, it being determined by the number of matrix microstates consistent with a given constant value of the trace Hamiltonian and of the Adler-Millard charge. The equilibrium distribution is obtained by maximising the entropy subject to the constancy of these two conserved quantities. Next, we must recall that the energy scale we are interested in is much smaller than Planck scale; as a consequence, the properties of the equilibrium distribution are determined by the AM charge, not by the trace Hamiltonian. A set of so-called Ward identities are proved for the equilibrium distribution, these being a generalised analog of the energy equipartition theorem in statistical mechanics. Important consequences follow from these identities. The AM charge is equipartitioned over all the bosonic and fermionic degrees of freedom and the equipartitioned value is identified with Planck’s constant. Recalling that the AM charge was defined in terms of the fundamental commutators and anti-commutators, it follows that the $q, p$ degrees of freedom, when averaged over the equilibrium canonical ensemble, obey the commutation relations of quantum theory. Lastly, it is shown that the canonically averaged configuration variables and momenta obey the Heisenberg equations of motion of quantum theory. This last result follows because the (canonical averages of the) functions which define the time derivatives in the (first-order) Hamilton’s equations of trace dynamics get related to the canonically averaged commutators which eventually appear in the Heisenberg equations of motion.

In this sense, quantum theory is an emergent phenomenon. When the equations of trace dynamics are coarse-grained over time intervals larger than Planck time, the emergent dynamics is quantum dynamics. This comes about because of the existence of the non-trivial Adler-Millard charge, unique to trace dynamics. The contact with quantum field theory is made by defining the Wightman functions of quantum field theory in terms of the emergent canonical averages of corresponding degrees of freedom in trace dynamics. One arrives at the standard relativistic quantum field theory for bosons and fermions. Since the Heisenberg equations of motion are now available, an equivalent Schrodinger picture dynamics can also be formulated.

Trace dynamics is one approach to derive quantum theory from symmetry principles, rather than arriving at quantum theory through the ad hoc recipe of `quantise the classical theory’. How can one justify the necessity of statistical mechanics in arriving at quantum theory? We recall that there is also an aspect of randomness/probabilities related to quantum mechanics - this being the aspect that comes into play during a quantum measurement. Randomness and probabilities are characteristic of statistical mechanics, specifically when fluctuations away from equilibrium become important. It is these fluctuations which are responsible for spontaneous localisation. We will take up this novel aspect of trace dynamics in the next section.

\section{The origin of spontaneous localisation: trace Dynamics (III)}

We have seen that the theory of trace dynamics gives rise to relativistic quantum (field) theory as an emergent phenomenon, after one constructs the equilibrium statistical thermodynamics of the underlying theory. This emergence opens up the possibility that spontaneous localisation is also a consequence of trace dynamics, in the following sense. When one arrives at the thermodynamic approximation by constructing the equilibrium configuration by applying statistical mechanics to the underlying microscopic theory, it is assumed that statistical  fluctuations away from equilibrium are negligible. Under certain circumstances though, fluctuations could become important. A situation of precisely this kind gives rise to spontaneous localisation, starting from the underlying trace dynamics, subject to certain assumptions. These assumptions will be justified when we incorporate gravity into trace dynamics (to be discussed in the next section).

Recall that the Adler-Millard charge is anti-self-adjoint, whereas the trace Hamiltonian is assumed to be self-adjoint.  We wish to consider the impact of statistical fluctuations on the emergent quantum dynamics. The description of these  fluctuations, and the stochastic correction that they imply to the trace Hamiltonian, is controlled by the Adler-Millard charge.  While the averaged Adler-Millard charge is equipartitioned, and proportional to Planck’s constant times a unit imaginary matrix, the fluctuations need not be equipartitioned. Moreover, one can by hand assume that the fluctuations can have an imaginary part, thus resulting in the Hamiltonian having an anti-self-adjoint part. [In the theory of spontaneous quantum gravity, this does not have to be put in by hand - the fundamental Hamiltonian at the Planck scale naturally has an anti-self-adjoint part].

The inclusion of imaginary fluctuations around equilibrium paves the way for spontaneous localisation to arise, provided some additional assumptions are made. It is assumed that such localisation takes place only for fermions (i.e. only in the matter sector, not for bosonic fields). Furthermore, only the non-relativistic case is considered, i.e. the Schrodinger equation for matter particles. Stochastic linear terms are added to the Hamiltonian, to represent fluctuations about equilibrium. Since the fluctuations include an imaginary part, the evolution of the modified Schrodinger equation does not preserve the norm of the state vector. [Norm preservation is essential, if one is to derive the Born probability rule]. The requirement of norm preservation is sought to be justified on empirical grounds, because particle number is conserved in the non-relativistic theory, and is related to the norm. Hence, a new state vector - which preserves norm - is defined by scaling the old state vector by its  norm. This makes the modified Schrodinger equation into a stochastic non-linear differential equation, which with the further assumption of no superluminal signalling, acquires precisely the form as in the GRWP theory of spontaneous collapse. And collapse does take place at a faster rate when more and more particles are entangled with each other. Entanglement enhances spontaneous collapse, making the equilibrium unstable. In fact, we have to recall that the mean dynamics arose after averaging over time scales larger than Planck time. Precisely this assumption - that such an averaging can be done - breaks down for macroscopic systems, as we will see in spontaneous quantum gravity! Moreover, there we will also find justification for the assumptions made above.

The take away note from this post is that trace dynamics contains within itself the roots for explaining spontaneous collapse, because quantum dynamics in the first place arises as a statistical thermodynamics approximation to the underlying matrix dynamics. There is every cause for investigating circumstances where fluctuations become important. And the classical world arises precisely because of such circumstances. Trace dynamics is presently the only known theory which provides a theoretical basis for spontaneous collapse.

A major limitation of trace dynamics is that while it operates at the Planck scale, it assumes space-time to  be Minkowski space-time. This however is clearly only a `transitory’ assumption, meant to be eventually relaxed. Until recently, it was not clear how to incorporate gravity as a matrix dynamics. We now know that Alain Connes’ non-commutative geometry programme enables us to do that. In so doing, we will also see how various assumptions made in trace dynamics get justified.

Incorporating gravity into trace dynamics finally helps us understand that spontaneous localisation arises because the fundamental Hamiltonian at  the Planck scale is not self-adjoint. It does not have to be. Only the emergent Hamiltonian in quantum theory has to be self-adjoint. We will also see how we arrive at a formulation of quantum theory which does not depend on classical space-time: this was one of our stated goals. We will also have a theory of quantum gravity which dynamically explains absence of superposition of classical space-time geometries. The theory also explains the origin of black hole entropy, and suggests a quantum gravitational origin for dark energy.

In the next section, we provide an overview of spontaneous quantum gravity.

\section{The theory of spontaneous quantum gravity: an overview}

As we have seen in the previous sections, we would like an underlying theory for spontaneous collapse, which also provides  a relativistic description of collapse. Trace dynamics goes a good part of the way, by deriving quantum field theory as the equilibrium statistical mechanics of an underlying matrix dynamics with global unitary invariance. Brownian motion fluctuations about equilibrium can provide the origin of spontaneous localisation, subject to a few assumptions. Trace dynamics operates at the Planck scale, but assumes space-time to be flat Minkowski space-time. Quantum theory is then derived by coarse graining trace dynamics over time scales much larger than Planck time. Quantum dynamics is thus a low energy emergent phenomenon, emerging after this coarse graining.

It is desirable though that we include gravity in trace dynamics, considering that Planck scale physics is involved. We also recall our other goal to have a formulation of quantum (field) theory without classical space-time. It turns out that the TD formalism can help us do that, if we can find a way to incorporate gravity. This would then also be a theory of quantum gravity. We also emphasised earlier that quantum gravity must dynamically explain the absence of space-time superpositions in the classical limit. That goal will be achieved here, because TD already has a mechanism (fluctuations) to explain spontaneous collapse.

So, how do we bring in gravity? It cannot be brought in simply as classical general relativity. That is not allowed by the Einstein hole argument: the matter in TD is not classical, whereas classical matter fields are needed to give operational meaning to the point structure of classical space-time. We also recall that trace dynamics describes matter degrees of freedom as matrices, which do not commute with each other. We search for a similar matrix/operator type description of gravity, which should not already be tied to quantisation of classical gravity. Because quantum theory should in fact emerge from the sought for [trace dynamics + gravity] after coarse-graining over Planck time scales. Thus we seek a matrix dynamics description of [matter + gravity], on the Planck scale.

Fortunately, such a matrix type description of gravity exists - it is the non-commutative geometry (NCG) program discovered and developed  by Alain Connes and collaborators. From our point of view, keeping trace dynamics in mind, NCG could be introduced as follows: what kind of geometry would we get if we raised space-time points (described by real numbers) to the status of matrices/operators? Recall we did just this kind of thing for material particles and gauge fields in TD. Similarly, our space-time points now become operators, which in general do not commute with each other.

In a mathematical approach, this can be understood as follows. Physical space, or space-time, is described by the laws of geometry. It can be mapped to an algebra, by assigning coordinates to the points of space. Then geometric properties (such as curvature) can be described in terms of functions on the algebra. This of course is a commutative algebra - real numbers commute with each other. After mapping the geometry of the space to a (commutative) algebra, we now take the following step: we make the algebra non-commutative. This is precisely what is achieved by elevating points of the space (or space-time) to matrices. The matrices do not commute with each other, and hence we have a non-commutative algebra. Now we ask: what kind of geometry such a non-commutative algebra describe?! That `geometry’, we call non-commutative geometry. There is no corresponding geometry in the sense in which we relate geometry to space, but we can talk of analogous concepts: e.g. what is the curvature of a non-commutative space?

One immediately notices the striking parallel between trace dynamics on the one hand, and NCG on the other. Both obtain by elevating classical point structures to the status of non-commuting operators. The former arrives at a matrix dynamics for matter and gauge fields. The latter arrives at a matrix description of geometry. Now classical general relativity couples classical matter to Riemannian geometry: matter curves space-time. We then expect matter described by trace dynamics to couple to non-commutative geometry: matrix matter `curves’ non-commutative space-time. Thus we intend to built a matrix theory of matter + gravity by unifying trace dynamics with non-commutative geometry. And we demand that this matrix dynamics have the following properties: when we perform the statistical mechanics of this gravity-based matrix dynamics, we should obtain a quantum theory of gravity at equilibrium. Fluctuations should become important for macroscopic systems (macroscopic to be defined) and spontaneous collapse should then come into play, giving rise to an emergent classical space-time and classical matter fields, such that the classical space-time has a Riemannian geometry which obeys the laws of general relativity. Fortunately, a mathematical formalism for such a programme has been developed: this is the theory of Spontaneous Quantum Gravity (SQG). SQG is in fact trace dynamics + trace gravity (i.e. NCG). From here, quantum gravity, quantum (field) theory, and classical general relativity, and classical dynamics, are all emergent phenomena.

What would such a matrix gravity look like, mathematically? Naively, we might want to make coordinates into operators, raise each metric component to an operator, try to construct a curvature tensor operator, and somehow couple it to the trace Lagrangian for matter fields in TD. But this does not work, for various reasons. It is technically difficult to make an invariant four-volume from the determinant of the metric, when each metric component is an operator. It all does not have the right feel, to say the least. Furthermore, classical space-time has been lost; how will we even describe time evolution and hence the dynamics, in matrix gravity?

Fortunately, the formalism of NCG shows the way forward. One can properly describe concepts such as distance and metric, in a non-commutative geometry. What is very important is that NCG seems to provide a new fundamental time parameter - a property unique to non-commutative geometry, not found in commutative algebras. We will return to this in some detail in future work - for now we just accept and employ this time parameter, which we will call Connes time. We lost space-time, but we recover time, and that is adequate for dynamics. Time is more fundamental than space.

When we raise space-time points to operators/matrices, like in TD, we can try to use the TD language of Grassmann matrices. In classical GR, metric is a field that lives on space-time. That won’t do now: trying to make something live on an operator. We expect the space-time operator to describe space-time geometry itself, and indeed that does happen. Also, we need to ask - the Grassmann matrix that represents space-time geometry: should it be just bosonic, or should it have a fermionic part too? I don’t at present know the answer to this important question. For now, we work with a Grassmann even (i.e. bosonic) matrix to describe spacetime geometry.

Since we want matrix gravity to yield GR (with matter sources) in the classical limit, we will have to specify a Lagrangian - both for gravity and for matter. Again, NCG shows the way, for gravity. There is a remarkable result in geometry, which relates curvature in a Riemannian geometry, to the Dirac operator on this space-time. Consider a Riemannian space - having a Euclidean signature. For now, and in this SQG program, we work with the Euclidean case. The Lorentzian case remains to be developed. Given a curved Riemannian space, one can write the standard Dirac operator $D_B$ on it 
[$i\gamma^\mu\ \partial_\mu$] in terms of the gamma-matrices and the spatial derivatives. A result from geometry states that [expressed for now as a simplified statement] the trace of the square of the Dirac operator is equal to the Einstein-Hilbert action [$\int d^4 x \ \sqrt{g} R$]. Isn’t that surprising - that the sum of the eigenvalues of the Dirac operator on a space is connected to Riemannian curvature on that space! The eigenvalues of the Dirac operator are connected to gravity - in fact they *are* gravity, as we will see later. The metric can be connected to these eigenvalues.

So we have this operator, $D_B^2$ on a Riemannian space. We can make the algebra of coordinates non-commutative, and we will still have this square of the Dirac operator: it describes curvature on a non-commutative space. And we have the Connes time, labelled say $\tau$, to describe evolution. We can now make contact with trace dynamics; recall that the trace Lagrangian is trace of an operator polynomial in configuration variables and their velocities. The trace polynomial coming from NCG is trace of square of the Dirac operator. Remembering that in quantum mechanics the Dirac operator is like momentum, we now introduce in our theory a bosonic operator/matrix $q_B$ such that its derivative with respect to Connes time is the Dirac operator $D_B$. This is the defining condition for $q_B$ while the defining condition for the Dirac operator is as before: it becomes the ordinary Dirac operator on a Riemannian space, and there it relates to the Ricci scalar and the Einstein-Hilbert action. In matrix gravity, the action describing gravity is the (Connes) time integral of the trace of the squared Dirac operator. This has just the form expected from trace dynamics. Moreover, the Lagrange equation resulting from this action is also very simple: the momentum is constant in time, and the configuration variable evolves linearly with time.

Next, we must include matter, because we after all want to derive spontaneous localisation of matter, from the SQG theory. At this stage in this programme we consider matter fermions only, leaving the consideration of (bosonic) gauge fields and non-gravitational interactions for later. So we have to have a way to include say Dirac fermions, in the language of trace dynamics. One thing we can anticipate is that these will be described by fermionic Grassmann matrices. But what should the Lagrangian be, keeping also in mind that we also have the Dirac operator at hand. We could construct a trace Lagrangian for every fermion in the theory, add up these Lagrangians, and add this to the trace Lagrangian for gravity (described above), integrate it over Connes time, and that could give the action for matrix dynamics.

However, we do not go on that path, for conceptual reasons. Let us ask the question: what is the gravitational  effect of an electron? An electron, being quantum mechanical, is all over space; so why must we distinguish the gravitational effect of the electron from the electron itself? This situation is unlike that of a classical object, say planet earth, where the object is localised in space, and its gravitational field is spread out everywhere. So we propose to introduce the concept of an `atom’ of space-time-matter [STM] which is a combined description of the fermionic part (say the electron) to be described by a fermionic operator $q_F$, and its gravitation part, to  be described by the bosonic $q_B$. Thus, we define the operator $q$ for an STM atom, written in terms of its bosonic and fermionic parts: $q = q_B + q_F$. For instance, in matrix gravity, an electron along with its gravity is an STM atom - it comes with its own operator space-time coordinates, its own Dirac operator. Further, we define the fermionic part of the Dirac operator, $D_F$ to be the Connes time derivative of $q_F$. The operator $D_F$ is determined by the requirement that in the classical limit it must give rise o the matter part of Einstein equations. 
 And the full Dirac operator $D$ is given by $D = D_B + D_F$. Recall that the original Dirac operator $D_B$ is bosonic. The Lagrangian for an STM atom is the trace of the square of $D$, and the action is the time integral of this Lagrangian. There is one such term for each STM atom. We can write the total action for matrix gravity simply as: 
\begin{equation}
S = \int d\tau \ \sum_i \ Tr [D_i^2]
\end{equation}

This is nice. From here we can derive quantum gravity, quantum theory, classical general relativity, all as emergent phenomena. To begin with, one easily obtains the Lagrange equations for the bosonic and fermionic part of each STM atom. The momenta are constant, and the configuration variables evolve linearly with time. Because of unitary invariance, there again is a conserved Adler-Millard charge. The figure below helps us understand where to go next.  
\begin{figure}[!htb]
        \center{\includegraphics[width=\textwidth]
        {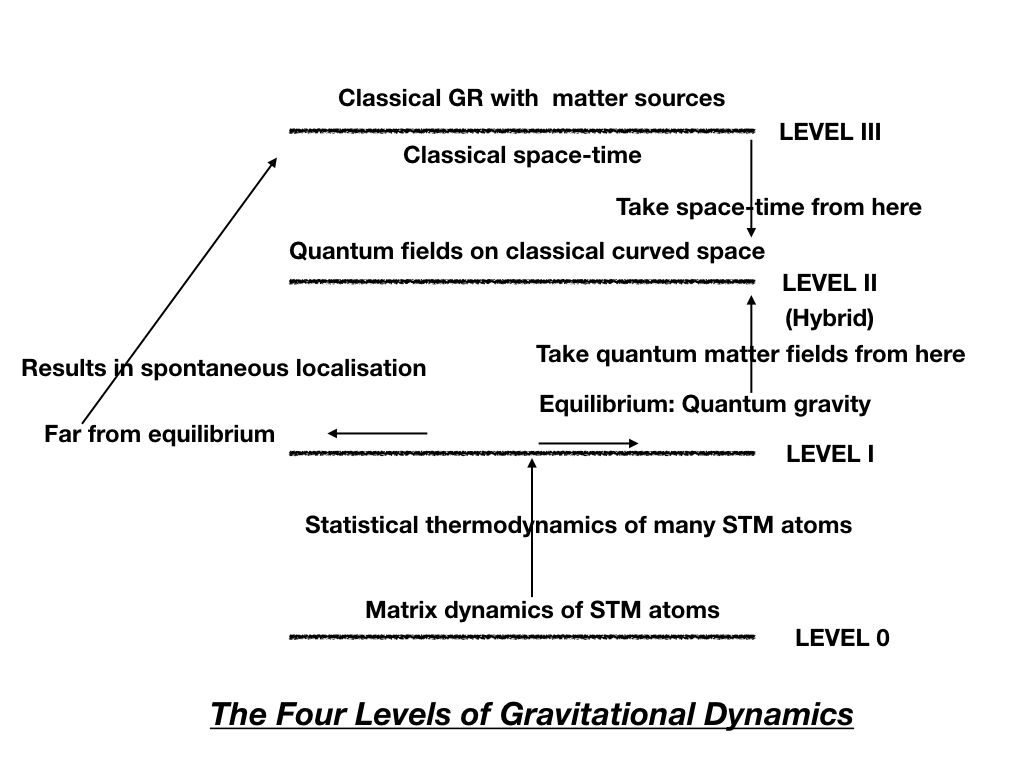}}
        \caption{\label{fig:my-label} The four levels of gravitational dynamics. In this bottom-up theory, the fundamental Level 0 describes the `classical' matrix dynamics of atoms of space-time-matter (STM). This level operates at the Planck scale. Statistical thermodynamics of these atoms brings us below Planck scale, to Level I: the emergent equilibrium theory is quantum gravity. Far from equilibrium, rapid spontaneous localisation results  in Level III: emergence of  classical space-time, obeying classical general relativity with matter sources. Level II is a hybrid level built by taking classical space-time from Level III and quantum matter fields from Level I, while neglecting the quantum gravitation of Level I. Strictly speaking, all quantum field dynamics takes place at Level I, but we approximate that to Level II. From \cite{maithresh2019b}.}
      \end{figure}
      
What we have described so far takes place at Level 0. There are only two fundamental constants at Level 0 - Planck length and Planck time. And there is the conserved Adler-Millard charge. Every STM atom has only one associated parameter, a length scale $L$, which eventually gets interpreted as Compton wavelength ($L$ can be different for every atom). At Level 0,  there is no Planck’s constant, no Newton’s gravitational constant, no concept of mass nor spin: all these are emergent at higher levels. At level 0, there is only the length scale $L$, from which mass and spin emerge subsequently at Level I. How do the STM atoms interact with each other? `Collisions’ and entanglement are possible mechanisms - this aspect is currently under investigation.

Level 0 is a Hilbert space on which the operators describing STM atoms live, and evolve in Connes time. Dynamics takes place at the Planck scale, as in trace dynamics. There is no space-time here. Space-time and the laws of general relativity emerge at Level III, as a consequence of spontaneous localisation. We can say that space-time arises as a consequence of collapse of the wave-function; more specifically, the part of the wave function that describes the fermions. The bosonic part does not undergo localisation, and becomes space-time and its curvature.

Like in trace dynamics, we would like to know what the emergent dynamics at low energies is, if we average Level 0 dynamics over time scales much larger than Planck time. For this we perform the statistical thermodynamics of the STM atoms described by the action given in the equation above. What emerges, at equilibrium at Level I, are the standard quantum commutation relations, and Heisenberg equations of motion, separately for the bosonic and fermionic parts of each STM atom. Evolution is still in Connes time. Planck’s constant emerges too, and hence Newton’s gravitational constant can be defined, using Planck’s constant along with Planck time and Planck length. The mass of an STM atom is defined in terms of its length $L$, which length hence can be interpreted as its Compton wavelength. The Schwarzschild radius of an STM atom is defined as square of Planck length divided by $L$. One can transform to the Schrodinger picture dynamics as well, and define quantum entanglement. Thus what we have at equilibrium at Level I is a quantum theory of gravity, emergent from the Level 0 matrix gravity dynamics. If we would like to know what is the gravitation of an electron, we can answer that question at Level I, or at Level 0, but not at Level II or III. Note that this Level I quantum gravity is a low energy phenomenon! It does not have anything to do with the Planck scale, but rather comes into play whenever a background space-time is not available. This Level I quantum gravity is also the sought after description of quantum (field) theory without classical time.

If a sufficiently large number of STM atoms get entangled, something very interesting takes place. If the total mass of the entangled system of STM atoms goes above Planck mass, the effective Compton wavelength of the full system goes below Planck length. The approximation that we can coarse grain the Level 0 dynamics over times larger than  Planck times breaks down. This is what we mean by fluctuations becoming important. The entangled system experiences rapid Planck scale fluctuations, an anti-self-adjoint part from the fermionic trace Hamiltonian is no longer negligible, and the entangled system undergoes extremely rapid spontaneous localisation. The localisation of the fermionic parts of many such entangled systems gives rise to the macroscopic bodies of the universe. Their bosonic parts together describe classical gravity, which is shown to obey the laws of classical general relativity.

Those STM atoms which do not undergo spontaneous localisation are to be described at Level 0 or Level I. Or, if we neglect their gravity, we can describe them at the hybrid Level II, after borrowing the space-time part from Level III. This is how we  conventionally do quantum (field) theory.

This theory of Spontaneous Quantum Gravity \cite{maithresh2019} is falsifiable, and makes the following predictions:

1. Spontaneous localisation (the GRW theory) is a prediction of this  theory, and the GRW theory is being tested in labs currently. If the GRW theory is ruled out by experiments, this proposal will be ruled out too.

2. SQG predicts the novel phenomena of quantum interference in time, and spontaneous collapse in time 
\cite{Singh:2019}.

3.The theory predicts the Karolyhazy length as a minimum length. This is testable and falsifiable \cite{SinghqgV2019}.

4. This theory predicts that dark energy is a quantum gravitational phenomenon \cite{Singh:DE}.

5. The theory provides an explanation for black hole entropy, from the microstates of STM atoms \cite{maithresh2019b}.

SQG is a candidate cover theory for general relativity, in the sense discussed in the first section. It explains the emergence of the classical world from quantum gravity, without having to resort to any interpretation of quantum mechanics.

The contents of this article are also available at qfqg.blogspot.com and will be developed further there.

\bigskip

\bigskip

\centerline{\bf REFERENCES}

\bibliography{biblioqmtstorsion}

\end{document}